\documentclass[twoside,12pt, a4paper]{article}
\usepackage{CJK}
\usepackage{indentfirst}
\usepackage{bm, amsmath, hyperref,xcolor}
\usepackage{epsfig}
\usepackage{subcaption}
\usepackage[labelformat=parens,labelsep=quad,skip=3pt]{caption}
\usepackage{graphicx}
\usepackage{cite}
\newcommand{\nn}{\nonumber}
\newcommand{\be}{\begin{equation}}
\newcommand{\ee}{\end{equation}}
\newcommand{\bea}{\begin{eqnarray}}
\newcommand{\eea}{\end{eqnarray}}

\usepackage{graphicx,cite}

\begin{document}

\begin{CJK*}{GBK}{song}


\begin{center}
\LARGE\bf Thermodynamics of soft wall model of AdS/QCD in Einstein-Maxwell-Gauss-Bonnet Gravity
\end{center}

\footnotetext{\hspace*{-.45cm}\footnotesize $^\dag$Corresponding author, E-mail: shobhitsachan@gmail.com, Presently at Shri Ramswaroop Memorial  University, Deva Road, Barabanki, India }

\begin{center}
\rm Shobhit Sachan$^{\rm a)\dagger}$, \ \ Sanjay Siwach$^{\rm b)}$, 
\end{center}

\begin{center}
\begin{footnotesize} \sl
${Department~ of ~Physics,~Banaras~ Hindu ~University,~Varanasi,~India}^{\rm a), b)}$ 
\end{footnotesize}
\end{center}
Email: shobhitsachan@gmail.com$^{\rm a)}$, sksiwach@hotmail.com$^{\rm b)}$


\vspace*{2mm}

\begin{center}
\begin{minipage}{15.5cm}
\parindent 20pt\footnotesize
\noindent
We investigate the thermodynamics of confinement/deconfinement transition in soft wall model of QCD with Gauss-Bonnet corrections using AdS/CFT correspondence. In bulk AdS space-time the transition is geometric and is known as Hawking-Page transition. The Hawking-Page transition between two geometries, namely charged AdS black hole and thermally charged AdS have been studied with Gauss-Bonnet corrections up-to first order. The Gauss-Bonnet coupling modifies the transition temperature of the system, but qualitative features remain unchanged. We obtain the curves between chemical potential and transition temperature for different values of Gauss-Bonnet couplings. We find that there exist a point in $\mu-T$ plane where lines with different value of Gauss-Bonnet coupling cross each other. This point may be the onset of the transition from first order to cross over behavior. The results are compared with that of the hard wall model.
\end{minipage}
\end{center}

\begin{center}
\begin{minipage}{15.5cm}
\begin{minipage}[t]{2.3cm}{\bf Keywords:}\end{minipage}
\begin{minipage}[t]{13.1cm}
AdS/CFT correspondence; Holographic QCD; Soft wall model.
\end{minipage}\par\vglue8pt

\end{minipage}
\end{center}

\section{Introduction}
\noindent
It has been a challenge to study the strongly coupled systems such as Quantum Chromodynamics (QCD) analytically. In particular, the low energy dynamics of QCD exhibit rich phase structure, but there is a very little hope that we can solve QCD in this regime by traditional perturbative methods. The lattice methods provide many important results about meson and baryon spectra and phase structure, but require extensive use of computational power. String inspired models have been developed recently and are used to study strongly correlated systems such as low energy QCD, high temperature superconductivity \cite{Sachdev:2010ch} etc.  Advent of AdS/CFT correspondence \cite{Maldacena:1997re,Witten:1998qj, Gubser:1998bc,Witten:1998zw} has motivated to look for super-gravity solutions \cite{Brandhuber:1998er} which can mimic the behaviour of realistic theories such as QCD. Some phenomenological models \cite{Erlich:2005qh, Da_Rold:2005zs} has been developed to understand the low energy dynamics of QCD and such an approach is dubbed as holographic QCD or AdS/QCD. String theory inspired models of QCD provide many interesting features of QCD phase diagram (see the recent reviews: \cite{Ammon:2015wua, CasalderreySolana:2011us, Nastase:2015a}).

The spectrum of mesons and baryons can be predicted by AdS/QCD approach \cite{Erlich:2005qh, Da_Rold:2005zs,deTeramond:2005su,Hong:2006ta,Erdmenger:2007cm} and seems to agree with experimental results. The dynamics of operators in QCD is captured by the equation of motion of the fields in bulk AdS space-time and leads to the realization of the chiral symmetry breaking, confinement/deconfinement transition \cite{Parnachev:2006dn,Da_Rold:2005zs,Cai:2007zw,BallonBayona:2007vp,Gherghetta:2009ac,Veschgini:2010ws,Megias:2010ku,Zhang:2010tk,Sachan:2011iy,Siwach:2014oya,Chelabi:2015gpc,Cao:2020ske,Chen:2018msc,Li:2017tdz,Fang:2016nfj,Chen:2019rez}, estimation of deconfinement temperature \cite{Katanaeva:2019anm}, pion condensation \cite{Lv:2018wfq}, quark-antiquark potential in deformed AdS/QCD models \cite{Bruni:2018dqm}. One can also study the various properties of the nuclear matter using the holographic models for AdS/QCD \cite{MartinContreras:2019kah,Vega:2018lym}.

The holographic QCD models can be studied using two approaches: top down models and bottom up models. The top-down approach of AdS/QCD is based on formalism developed in \cite{Klebanov:2000hb, Antonyan:2006vw, Babington:2003vm} and QCD like theories are constructed from various $D$ brane configurations. The first of such models is proposed by Karch et. al. \cite{Karch:2002sh}. They considered $D_3-D_7$ brane configuration. The most of the features of this model are studied in probe limit, where back-reaction of the flavor $D$ branes can be safely neglected. The incorporation of flavors in the model is done by space-time filling $D$ branes. The most successful and well established model  based on top-down approach is Sakai-Sugimoto model \cite{Sakai:2004cn, Sakai:2005yt}, in which the holographic dual large $N_c$ QCD with group structure $U(N_c)$ and number of flavors $N_f$ is realized by the probe $D_8$ branes in background of $D_4$ branes (see \cite{Rebhan:2014rxa} for a recent review).  

In  bottom up approach, also known as potential reconstruction approach \cite{He:2011hw, Li:2011hp} one starts from well established features of QCD and tries to formulate higher dimensional dual models (holographic models) which are capable of reproducing known results in QCD.  The potential reconstruction approach is used to study various properties of Einstein-Maxwell-Dilaton (EMD) system \cite{He:2013qq, He:2011hw}.  A lot of progress is still needed to construct  a full dual model which predicts most of the features of the QCD. The bottom-up approach is further divided into two categories: hard wall model and soft wall model. In the simplest model of AdS/QCD, one assumes that the theory can be described by the five dimensional model living on the slice of the $AdS_5$ or some deformations of it \cite{Erlich:2005qh,Da_Rold:2005zs,Andreev:2010bv}. This model, known as hard wall model of QCD  \cite{Erlich:2005qh} contains two gauge fields $A_{L\mu}^a$ and  $A_{R\mu}^a$, which are dual to left handed and right handed chiral symmetry currents and a bifundamental tachyonic field $X^{\alpha\beta}$. The tachyonic field breaks the chiral symmetry and plays the role of quark-antiquark condensate in boundary theory. The phenomenon of quark confinement is realized by the introduction of a fixed IR boundary and this boundary is identified with the inverse of the QCD scale $\Lambda$.

The hard wall model is widely used to predicts the excited state hadronic masses, coupling constants and decay constants of fields etc. and the predictions seem to agree with experimental results. However, the mass spectrum for $n^\mathrm{th}$ excited state as predicted by hard wall model varies as $m_n^2\propto n^2$, but the  QCD data requires that $m_n^2\propto n$. This requirement (Regge trajectories) lead to the development of soft wall model of QCD \cite{Karch:2006pv}. As compared to fixed IR boundary in hard wall model, the soft wall model has dynamical IR boundary. This flexibility in IR boundary is achieved by the introduction of dilaton field in the bulk action.

In this paper, we study the phase structure of QCD in soft wall model with Gauss-Bonnet  terms \cite{Cai:2001dz,Cvetic:2001bk} in the bulk theory. The Gauss-Bonnet term is simplest higher derivative corrections to Einstein's general relativity that leads to second order field equations for the metric. In string inspired phenomenological models the coupling constant $\alpha=1/4\pi\alpha'>0$ is related to stringy corrections to general relativity \cite{Myers:1987yn, Wheeler:1985nh, Cho:2002hq, Torii:2005xu}, where $\alpha'$ is the  string tension. In this sense, the inclusion of Gauss-Bonnet term in our study leads to stringy corrections to the Einstein-Maxwell system. This is the key motivation of our work.    

We consider the charged black hole solution of five dimensional Einstein-Maxwell theory with Gauss-Bonnet term and calculate the grand potential in this geometry. Gauss-Bonnet AdS black hole solution describes the high temperature quark-gluon plasma phase of the holographic dual QCD. The low temperature confined phase is described by thermally charged Gauss-Bonnet AdS geometry. We regularize both the actions by subtracting the thermal AdS action. The difference of grand potentials in these two geometries is used to obtain the transition temperature versus chemical potential graph of the soft wall model for different values of Gauss-Bonnet coupling.  

This paper is organized as follows. In the  next section, we consider the black hole solutions in  Einstein-Maxwell theory with Gauss-Bonnet corrections. The AdS black hole and thermally charged AdS solutions of the soft wall holographic model QCD are discussed in subsequent sections. The discussion of confinement/deconfinement transition is done in section 5, while the results are summarized in the last section.

\section{Gravitational  Solutions}
\noindent
Consider the Euclidean action of Einstein-Maxwell-Gauss-Bonnet theory with negative cosmological constant in five dimension; 
\begin{equation}\label{action0}
S~=~-\int~d^5x~\sqrt{g}~\left\lbrace\frac{1}{2\kappa^2}\left(R-2\Lambda+\alpha
R_{GB}\right)-\frac{1}{4g^2}F^2\right\rbrace,
\end{equation}
where, $R_{GB}=R^2-4R_{MN}R^{MN}+R^{MNPQ}R_{MNPQ}$ is the Gauss-Bonnet term and $F^2=F_{MN}F^{MN}$ is the field strength of the Maxwell field. The constant $\kappa^2$ is related to five dimensional Newton's constant $G_5$ through relation $\kappa^2=8\pi G_5$ and the cosmological constant is taken as $\Lambda=-6/L^2$. The Gauss-Bonnet corrections are the simplest form of higher derivative corrections to Einstein-Hilbert action. The more general contribution arises from Lanczos-Lovelock gravity \cite{Lovelock:1971yv, Padmanabhan:2013xyr}. The Gauss-Bonnet terms make contributions in the bulk only for dimensions higher than four. In space-time dimension dimension less than or equal to four, these terms behave as topological terms.

The Euler-Lagrange equations of motion with second order derivatives with Lagrangian density ${\cal L}$ are given by,
\begin{equation}
\frac{\partial {\cal L}}{\partial
g_{MN}}-\partial_P\left[\frac{\partial {\cal
L}}{\partial_P(\partial
g_{MN})}\right]+\partial_P\partial_Q\left[\frac{\partial {\cal
L}}{\partial_P\partial_Q(\partial g_{MN})}\right]~=~0.
\end{equation}
This general equation of motion leads to gravitational equations of motion for our Lagrangian as,
\begin{equation}\label{eom}
G_{MN}+\alpha H_{MN}=\frac{\kappa^2}{g^2}T_{MN}
\end{equation}
where the various terms used in above equation are given by,
\begin{subequations}
\begin{align}
H_{MN}~&=~2(R
R_{MN}-2R_{MP}R^{~~P}_N-2R^{PQ}R_{MPNQ}+R_M^{~~PQR}R_{NPQR})-\frac{1}{2}g_{MN}R_{GB}\\
G_{MN}~&=~R_{MN}-\frac{1}{2}R~g_{MN}+\Lambda~g_{MN}\\
T_{MN}~&=~F_{PM}F^P_{~~N}-\frac{1}{4}g_{MN}F^{PQ}F_{PQ}.
\end{align}
\end{subequations}
The variation of gauge fields leads to the Maxwell's equations of motion in curved space, which are written as,
\begin{equation}\label{gauge}
\frac{1}{\sqrt{g}}\partial_M(\sqrt{g}\,F^{MN})~=~0.
\end{equation}

The theory admits charged black hole solution with AdS asymptotics \cite{Cvetic:2001bk, Cai:2001dz}. Let us consider the Euclidean metric ansatz of the Gauss-Bonnet black hole as;
\be\label{metric1}
ds^2~=~\frac{L^2}{z^2}\left(A^2
f(z)\,dt^2+\frac{dz^2}{f(z)}+\sum_{i=1}^{3}{dx^i}^2\right)
\ee
where $A^2= \frac{1}{2} \left(\sqrt{1-8 \alpha }+1\right)$. The value of $A$ is fixed in such a way that it can lead to conformally flat metric at spatial infinity. 

The solution of the equation of motion of Einstein-Maxwell-Gauss-Bonnet (EMGB) theory leads to the identification of form of metric function $f(z)$ and is given by,
\be\label{metricfunction1}
f_1(z)~=~\frac{1}{4\alpha}(1-\sqrt{1-8\alpha(1-mz^4+q^2z^6)}~),
\ee
which, in the limit $\alpha\to 0$ leads to the metric of AdS RN black hole. 

We consider only the temporal component of gauge field and  the solution of gauge field equation is given by, 
\be
{\cal A}_t(z)~=~i (\mu-Q z^2),
\ee
where $i$ in front of solution is due to consideration of Euclidean (Wick's rotation) geometry in the study. The constant $\mu$ will be identified
as quark chemical potential in the dual QCD model and $Q$ is identified as quark number density and  is related to black hole charge.  

Contracting equation \eqref{eom} with $g^{MN}$, we get
\begin{equation}
-\frac{3}{2}R+5\Lambda-\frac{\alpha}{2}R_{GB}~=~-\frac{\kappa^2}{4g^2}F_{ab}F^{ab}.
\end{equation}
Using the above equation and \eqref{action0}, we get the simplified  form of action as,
\begin{equation}\label{action}
S~=~-\int~d^5x~\sqrt{g}~\left(\frac{-R+4\Lambda}{\kappa^2}\right).
\end{equation}
This equation shows that the dependence of of the action on Gauss-Bonnet coupling, $\alpha$  is not explicit but through the metric function and Ricci scalar.

To satisfy Einstein's and Maxwell's equation of motion,  the relation between black hole charge ($q$) and quark number density ($Q$) in EMGB gravity must be written as,
\begin{equation}\label{charge}
q^2~=~\frac{2}{3}\frac{\kappa^2}{g^2}\frac{Q^2}{A^2}.
\end{equation}
The AdS/QCD correspondence relates five dimensional gravitational constant $(\kappa^2)$ and five dimensional coupling
constants $(g^2)$ to the rank of colour gauge group ($N_c$) and number of flavours ($N_f$);
\begin{equation}
\frac{1}{2\kappa^2}~=~\frac{N_c^2}{8\pi^2}~~\textrm{and}~~\frac{1}{2g^2}~=~\frac{N_cN_f}{8\pi^2}.
\end{equation}

Let us fix the outer horizon at $z_+$, the metric function  $f_1(z)$ must satisfy the relation $f_1(z_+)=0$. This relation leads us to replace the black hole mass $m$ in terms of outer horizon $z_+$  and black hole charge $q$;
\begin{equation}
m=\frac{1}{z_+^4}+q^2~z_+^2.
\end{equation}
Using this relation, one can write the  Hawking temperature of black hole with Gauss-Bonnet corrections  as,
\begin{equation}\label{hwt}
T~=~\frac{A\,f_1'(z_+)}{4\pi}~=~\frac{A}{\pi~z_+}(1-\frac{1}{2}q^2~z_+^6).
\end{equation}
Since the solutions of the gauge field  ${\cal A}_t(z)$ are regular at the horizon, we impose Dirichlet boundary condition at the horizon $z_+$ as ${\cal A}_t(z_+)=0$. This leads to the relation,
\be\label{charge1}
q^2=2\kappa^2\mu^2/3g^2z_+^4A^2.
\ee
Now combining \autoref{hwt} and \autoref{charge1}, one can get a quadratic equation for $z_+$ ,
\be
\frac{1}{3\,A}\frac{\kappa^2}{g^2}\mu^2\,z_+^2+\pi\,\,T~z_+-A~=~0,
 \ee
which gives the (positive) solution for outer horizon, $z_+$ as,
\begin{equation}\label{horizon}
z_+~=~ \frac{3 A\, g^2}{2 \kappa ^2 \mu ^2} \left(\sqrt{\frac{4
}{3}\frac{\kappa ^2 \mu ^2}{g^2}+\pi ^2 T^2}-\pi T\right).
\end{equation}
This equation shows that the  horizon radius depends upon Gauss-Bonnet coupling $\alpha$ through $A$ only.

\section{Charged Black hole Solution}\label{sec:cb}
\noindent
The soft wall model of QCD was developed by A. Karch et. al. \cite{Karch:2006pv}. The phase structure of soft wall model  is investigated in \cite{Herzog:2006ra, Sachan:2011iy, Park:2011qq} and  give predictions about deconfinement temperature. The soft wall model contains an exponential dilaton field with IR boundary at infinity. The simplified action in soft wall model with Gauss-Bonnet corrections is written as,
\be\label{action1}
S~=~-\int~d^5x~\sqrt{g}~e^\phi\left(\frac{-R+4\Lambda}{\kappa^2}\right).
\ee
where $\phi$ is dilaton field given by the relation  $\phi(z)=-c\,z^2$, and the value of constant $c$ is soft wall model parameter and it is calculated by matching the substituting the mass of lightest meson in the solutions of equation of motion in soft wall model. Its value is estimated to be $\sqrt{c}=388~\mathrm{MeV}$ \cite{Herzog:2006ra}.  The dilaton field have nontrivial expectation values and is assumed not to affect the gravitational dynamics of the theory.

The action in the high temperature deconfined phase is given by black hole solution of the EMGB theory;
\be\label{action3}
S~=~-\int~d^5x~\sqrt{g}~e^{-c\,z^2}\left(\frac{z^2 f''(z)-8 z f'(z)+20 f(z)+4\Lambda}{\kappa^2}\right).
\ee
Since the integrals in the above action can not be done analytically, we expand the action up-to first order in Gauss-Bonnet coupling, $\alpha$ (and hence our results will be valid for small values of $\alpha$). The action up to first order in $\alpha$ is written as,
\begin{align}
S_1~&=~-\frac{A}{\kappa^2}\int d^5x~\frac{e^{-c\,z^2}}{z^5}\big[\left(2 q^2 z^6-4\right)+\alpha\left(24 m^2 z^8-120 m q^2 z^{10}+112 q^4 z^{12}+8 q^2z^6+40\right)\big]\nn\\
&=~-\frac{A}{\kappa^2}V_3\int_0 ^{\beta_1} dt\int_\epsilon^{z_+}dz~\frac{e^{-c\,z^2}}{z^5}\Big[\left(2 q^2 z^6-4\right)+\alpha \left(24 m^2 z^8-120 m q^2 z^{10}+112 q^4 z^{12}\right.\nn\\
&~~~~~~+\left.8 q^2 z^6+40\right)\Big]
\end{align}
where $\beta_1$ is periodicity of Euclidean space-time and defined as inverse of Hawking temperature $T$ and $V_3$ is the three dimensional volume. The lower limit is UV cut-off and we shall be taking the limit as $\epsilon\to 0$ at the end of the calculation. Evaluation of various integrals gives us the action as,
\be \label{action2}
S_1~=~\frac{A~V_3}{\kappa^2}\beta_1\big[(1-10\alpha)F_1(z)+(1+4\alpha)F_2(z)+\alpha F_3(z)\big]\Big\vert_\epsilon^{z_+},
\ee
where, the functions  $F_1,\,F_2\,\textrm{and}\,F_3$ are defined as,
\begin{align}
F_1(z)~=&~\frac{e^{-c\, z^2}}{z^4}(c\,z^2-1)+c^2\,\text{Ei}\left(-c\, z^2\right)\nn\\
F_2(z)~=&~\frac{e^{-c\, z^2}}{c}q^2\nn\\
F_3(z)~=&~e^{-c\,z^2}\left[12\,m^2\left(\frac{1}{c^2}+\frac{z^2}{c}\right)-60\,m\,q^2\left(\frac{2}{c^3}+\frac{2z^2}{c^2}+\frac{z^4}{c}\right)\right.\nn\\
&~~~~~~~~+\left.56\,q^4\left(\frac{6}{c^4}+\frac{6z^2}{c^3}+\frac{3z^4}{c^2}+\frac{z^6}{c}\right)\right]
\end{align}
and the exponential integral, Ei is given by $\textrm{Ei}(x)=-\int_{-x}^\infty dt\,e^{-t}/t$.

The first term of $F_1$ is divergent in the limit $z\to0$. This divergence will be removed by the subtraction of thermal AdS action from that of charged AdS black hole. The thermal AdS solution of Gauss-Bonnet gravity has the metric function as, 
\be
f_0~=~\frac{1}{4\alpha}(1-\sqrt{1-8\alpha}).
\ee
The form of the metric remains unchanged, but $f(z)$ in \eqref{metric1} is replaced by $f_0(z)$. $\alpha\to0$ limit of this metric corresponds to Euclidean AdS metric.
Using this metric function, the action up to order $\alpha$ for thermal AdS is given by,
\begin{align}
S_0~&=~-\frac{A}{\kappa^2}\int d^5x \frac{(40 \,\alpha -4) e^{-c \,z^2}}{z^5}~=~-\frac{A}{\kappa^2}V_3\int_0^{\beta_0}\int_\epsilon^\infty\frac{(40 \,\alpha -4) e^{-c \,z^2}}{z^5}\nn\\
&=~\frac{A}{\kappa^2}\beta_0V_3(1-10\,\alpha)F_1(z)\Big\vert_\epsilon^\infty,
\end{align}
where $\beta_0$ is time periodicity of thermal AdS and $\lim_{x\to-\infty}F_1(x)$ is zero since exponential integral Ei($x$) is zero in the limit $x\to-\infty$.The regularized action for charged AdS black hole with Gauss-Bonnet term is given by,
\begin{align}\label{eq:action4}
\bar{S}_1~&=~\lim_{\epsilon\to0}\left(S_1-S_0\right)\nn\\
&~=\frac{AV_3}{\kappa^2}\beta_1\Big[
(1-10\alpha)F_1(z_+)+(1+4\alpha) F_2(z_+)+\alpha F_3(z_+)\nn\\
&~~~~-(1+4\alpha)\, F_2(0)-\alpha\, F_3(0)+\frac{m}{2}-4m\,\alpha\Big],
\end{align}
where periodicity of two geometries are matched at $z=\epsilon$ using the relation,
\be
\beta_0~=~\beta_1\sqrt{\frac{f_1(\epsilon)}{f_0(\epsilon)}}.
\ee 
The last two terms in the expression \eqref{eq:action4} arises from subtraction of $F_1$ terms in both the geometries and we have considered terms up to order $\alpha$ in the regularization. Taking limit $\alpha\to0$, we recover the results of \cite{Park:2011qq}. 

The  grand potential and on-shell action are related by the thermodynamic relation, $\Omega=T\,S_{onshell}$. Thus the difference of grand potentials per unit volume in the two geometries is given by,
\begin{align}
\frac{\Omega_1}{V_3}~&=~\frac{A}{\kappa^2}\Big[
(1-10\alpha)F_1(z_+)+(1+4\alpha) F_2(z_+)+\alpha F_3(z_+)\nn\\
&-(1+4\alpha)\, F_2(0)-\alpha\, F_3(0)+\frac{m}{2}-4m\,\alpha\Big].
\end{align}

\section{Thermally charged AdS Solution}\label{sec:tc}
\noindent
The deconfinement phase is high temperature phase and is represented by charged AdS black hole described above, while the low temperature phase is confinement phase or hadronic phase and is represented by thermally charged AdS solutions\cite{Park:2011qq, Lee:2009bya}. The metric for thermally charged AdS with Gauss-Bonnet correction terms is written as,
  \begin{align}
   ds^2~&=~\frac{L^2}{z^2}\left(A^2 f_2(z)\,dt^2+\frac{dz^2}{f_2(z)}+\sum_{i=1}^{3}{dx^i}^2\right)\\\nn
  & \textrm{with}~f_2(z)=\frac{1}{4\alpha}\left(1-\sqrt{1-8\alpha(1+q_2^2z^6)}\right).
  \end{align}
The above metric in the limit $\alpha\to0$, corresponds to metric function of thermally charged AdS as described by Lee et. al. in \cite{Lee:2009bya} which is  asymptotically AdS. The thermal AdS lacks the horizon, therefore, one need a UV cut-off at $z\to 0$. 

The action for thermally charged AdS up to first order in $\alpha$ is given by,
\begin{align}
S_2~&=~-\frac{A}{\kappa^2}\int d^5x~\frac{e^{-c\,z^2}}{z^5}\big[\left(2 q_2^2 z^6-4\right)+\alpha\left(112 q_2^4 z^{12}+8 q_2^2z^6+40\right)\big]\nn\\
&=~-\frac{A}{\kappa^2}V_3\int_0 ^{\beta_2} dt\int_\epsilon^{\infty}dz~\frac{e^{-c\,z^2}}{z^5}\Big[\left(2 q_2^2 z^6-4\right)+\alpha \left(112 q_2^4 z^{12}+8 q_2^2 z^6+40\right)\Big],
\end{align}
where $\beta_2$ is time periodicity of thermally charged AdS. Evaluation of this action and rearranging the various terms, the final action is of form,
\be \label{action4}
S_2~=~\frac{A~V_3}{\kappa^2}\beta_2\big[(1-10\alpha)F_1(z)+(1+4\alpha)F_4(z)+\alpha F_5(z)\big]\Big\vert_\epsilon^{\infty}
\ee
where,
\be
F_4(z)~=~\frac{e^{-c\, z^2}}{c}q_2^2, \qquad\textrm{and}\qquad F_5(z)~=~56\,e^{-c\,z^2}\,q_2^4\left(\frac{6}{c^4}+\frac{6z^2}{c^3}+\frac{3z^4}{c^2}+\frac{z^6}{c}\right).
\ee
The regularized action for thermally charged AdS is calculated by subtraction of thermal AdS action with Gauss-Bonnet coupling and given by,
\begin{align}
\bar{S}_2~&=~\lim_{\epsilon\to0}\left(S_2-S_0\right)\nn\\
&=~\frac{A~V_3}{\kappa^2}\beta_2\big[-(1+4\alpha)\, F_4(0)-\alpha\, F_5(0)\big],
\end{align}
where time periodicity of thermal AdS and thermally charged AdS are related as $\beta_0=\beta_2\sqrt{\frac{f_2(\epsilon)}{f_0(\epsilon)}}$. 

The the grand potential per unit volume in this geometry is given by,
\be
\frac{\Omega_2}{V_3}~=~-\frac{A}{\kappa^2}\big[(1+4\alpha)\, F_4(0)+\alpha\, F_5(0)\big]~=~-\frac{A}{\kappa^2}\left[\frac{q_2^2}{c}+4\alpha\left(\frac{q_2^2}{c}+\frac{84~q_2^4}{c^4}\right)\right].
 \ee 
\begin{figure}
  \begin{subfigure}[b]{0.5\linewidth}
    \centering
    \includegraphics[width=8cm]{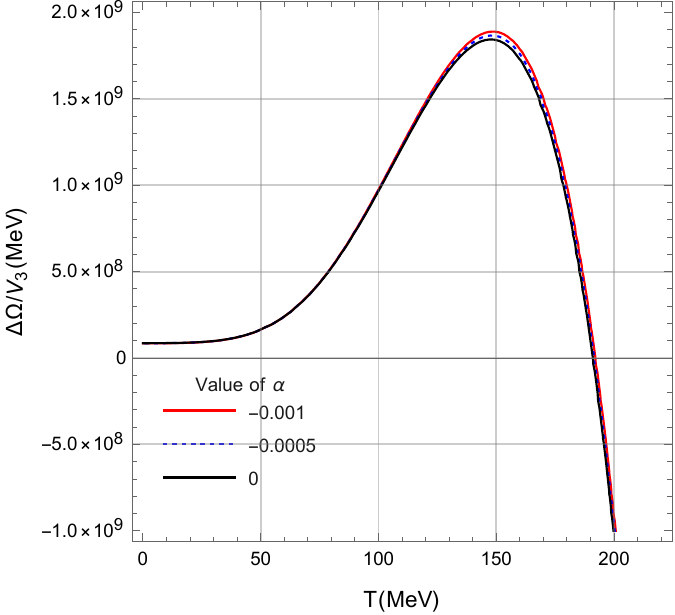}
    \caption{For $\mu=50~MeV$.}
  \end{subfigure}\hfill
  \begin{subfigure}[b]{0.5\linewidth}
    \centering
    \includegraphics[width=8cm]{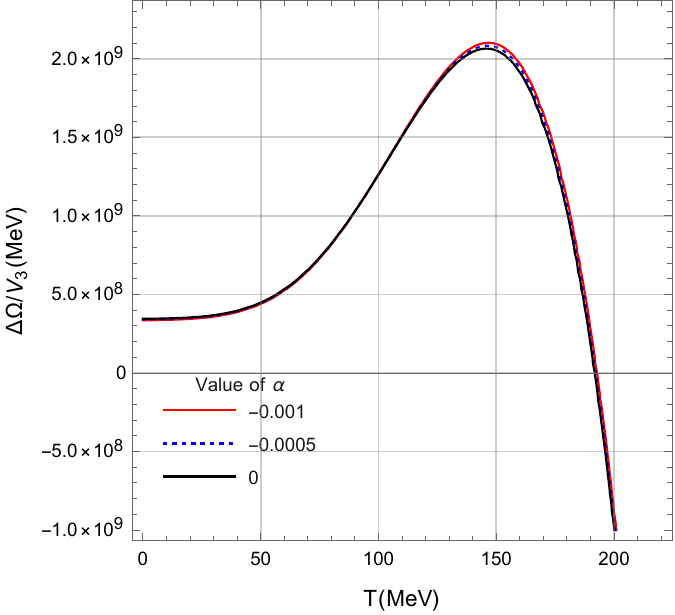}
    \caption{For $\mu=100~MeV$.}
  \end{subfigure}
  
  \begin{subfigure}[b]{0.5\linewidth}
    \centering
    \includegraphics[width=8cm]{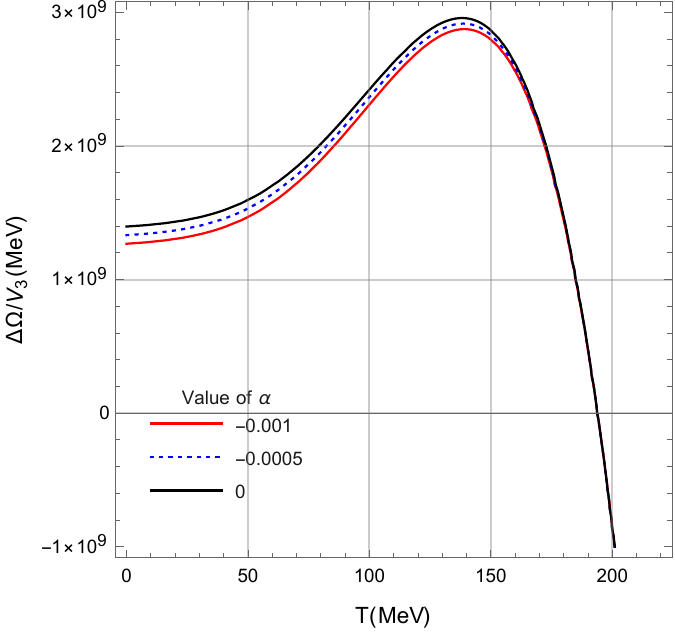}
    \caption{For $\mu=200~MeV$.}
  \end{subfigure}\hfill   
    \begin{subfigure}[b]{0.5\linewidth}
      \centering
      \includegraphics[width=8cm]{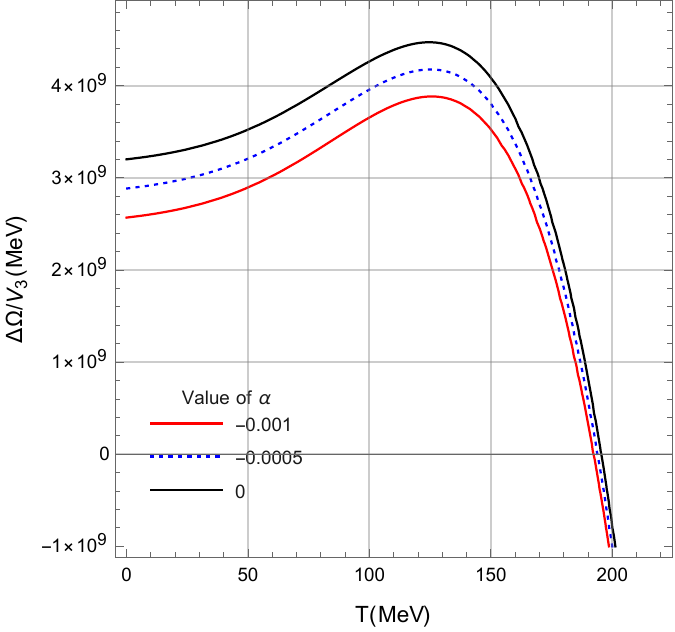}
    \caption{For $\mu=300~MeV$.}
  \end{subfigure}
  
   \begin{subfigure}[b]{0.5\linewidth}
    \centering
    \includegraphics[width=8cm]{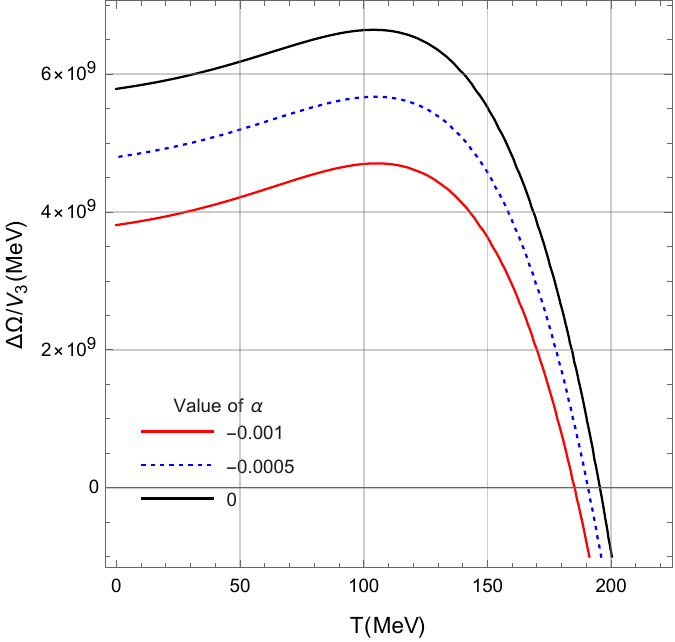}
    \caption{For $\mu=400~MeV$.}
  \end{subfigure}\hfill
  \begin{subfigure}[b]{0.5\linewidth}
    \centering
    \includegraphics[width=8cm]{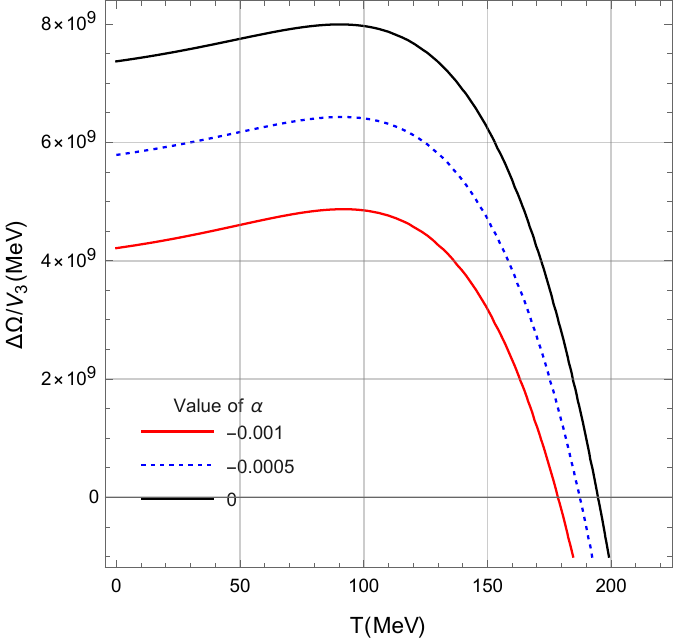}
    \caption{For $\mu=450~MeV$.}
  \end{subfigure}
  \caption{Grand potential difference vs temperature for different values of Gauss-Bonnet coupling in soft wall model.}\label{fig:GB_free}
\end{figure}

The grand potential is function of $\mu$, therefore $q_2$ must be function of $\mu$. Let  $q_2$ and  $\mu$ are related linearly and the relation is $q_2=i \zeta \mu$, where $\zeta$ is unknown constant to be determined by boundary conditions. This assumption of linearity between $q_2$ and $\mu$ is for zeroth order in $\alpha$. Using this assumption, we equate the boundary value of action$^\ddagger$\footnote{$^\ddagger$using Dirichlet boundary conditions at UV boundary gives free energy from Legendre transformation of grand potential  \cite{Park:2011qq, Lee:2009bya}.} to $\mu N \beta_2$, where $N$ is number of quarks given by relation $N=-\partial\Omega_2/\partial\mu$. Thus the value of $q_2$ is given by equation, 
\be
q_2~=~\sqrt{\frac{3}{2}}~\frac{\kappa}{g}~\frac{c\,\mu}{A}.
  \ee
  
\section{Confinement/deconfinement Transition}\label{sec:tra1}
\noindent
The confinement/deconfinement transition can be studied in the soft wall model with Gauss-Bonnet corrections. We consider the leading order terms in Gauss-Bonnet coupling as the integrals involving soft model are intractable analytically. In the bulk five dimensional space-time this transition is geometric and known as Hawking-Page  transition \cite{Hawking:1982dh}. The Hawking-Page transition is gravitational dual of confinement/deconfinement transition within the finite volume boundary theory. The higher temperature deconfined phase is realized by the black hole geometry and low temperature confined phase is captured by thermal AdS space-time. Here, we study the dependence of difference in grand potential in these two geometries and investigate the effects of  the Gauss-Bonnet coupling $\alpha$ on the transition temperature.
   
Using the computation of Euclidean action of the soft wall model in two geometries,  the difference in grand potential is given by,
\begin{align}
\frac{\Delta\Omega}{V_3}~&=~\frac{\Omega_1-\Omega_2}{V_3}\nn\\
&=~\frac{A}{\kappa^2}\left[
(1-10\alpha)F_1(z_+)+(1+4\alpha) F_2(z_+)+\alpha F_3(z_+)-(1+4\alpha)\, F_2(0)-\alpha\, F_3(0)\right.\nn\\
&~~~~\left.+\frac{m}{2}-4m\,\alpha+(1+4\alpha)\, F_4(0)+\alpha\, F_5(0)\right]
\label{freediff}
\end{align}
where various functions  are summarized as,
\begin{align}
F_1(z)~=&~\frac{e^{-c\, z^2}}{z^4}(c\,z^2-1)+c^2\,\text{Ei}\left(-c\, z^2\right)\nn\\
F_2(z)~=&~\frac{e^{-c\, z^2}}{c}q^2\nn\\
F_3(z)~=&~e^{-c\,z^2}\left[12\,m^2\left(\frac{1}{c^2}+\frac{z^2}{c}\right)-60\,m\,q^2\left(\frac{2}{c^3}+\frac{2z^2}{c^2}+\frac{z^4}{c}\right)+56\,q^4\left(\frac{6}{c^4}+\frac{6z^2}{c^3}+\frac{3z^4}{c^2}+\frac{z^6}{c}\right)\right]\nn\\
F_4(z)~=&~\frac{e^{-c\, z^2}}{c}q_2^2\nn\\
F_5(z)~=&~56\,e^{-c\,z^2}\,q_2^4\left(\frac{6}{c^4}+\frac{6z^2}{c^3}+\frac{3z^4}{c^2}+\frac{z^6}{c}\right).\nn
\end{align}
\begin{figure}
  \begin{subfigure}[b]{0.5\linewidth}
    \centering
    \includegraphics[width=8cm]{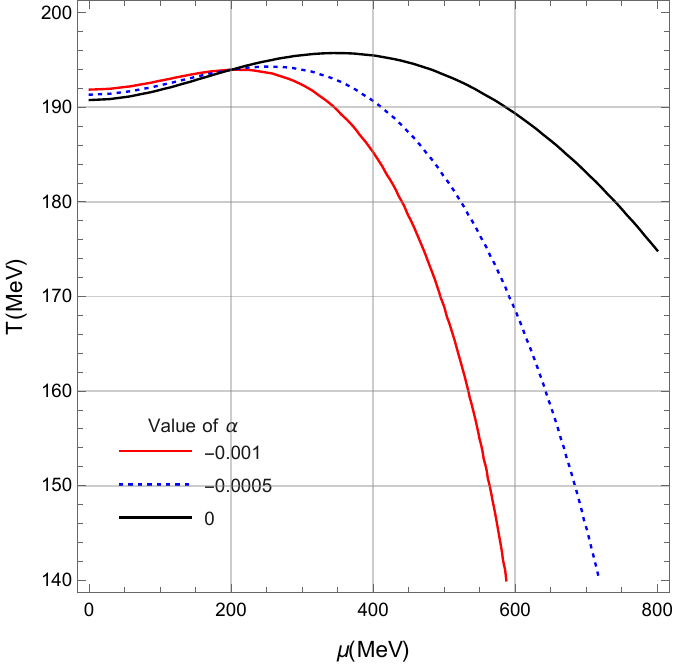}
    \caption{Soft wall model.}\label{fig:GB_tmu1}
  \end{subfigure}\hfill
  \begin{subfigure}[b]{0.5\linewidth}
    \centering
    \includegraphics[width=8cm]{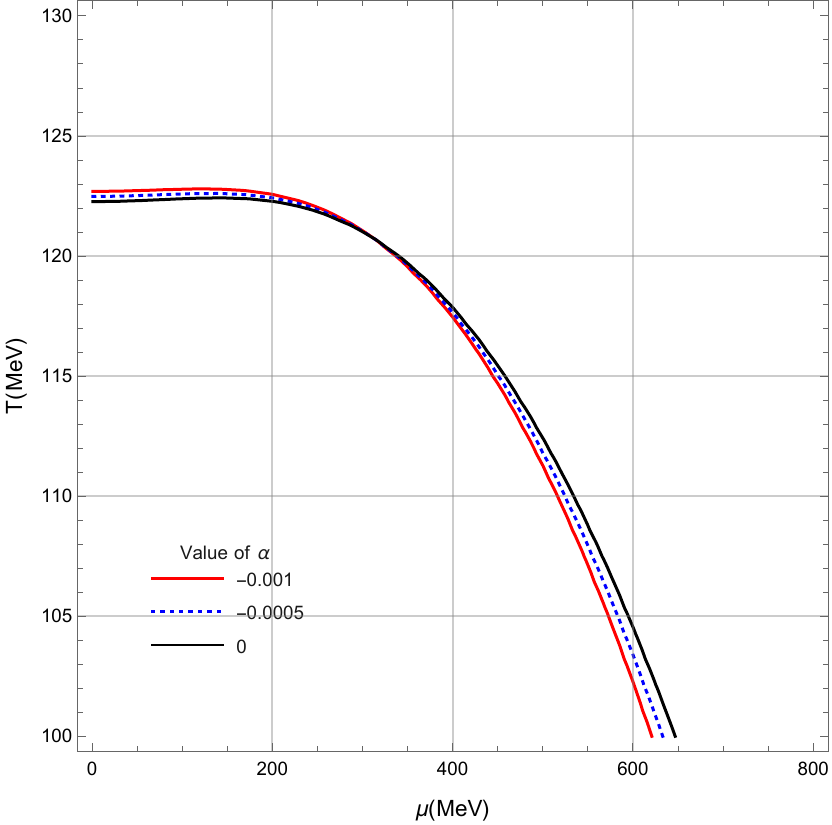}
    \caption{Hard wall model.}\label{fig:hw_tmu}
  \end{subfigure}
  \caption{Transition temperature vs chemical potential for different values of Gauss-Bonnet coupling $\alpha$.}\label{fig:GB_tmu}
\end{figure}
The charges in both the geometries are given by relations,
\[q~=~\sqrt{\frac{2}{3}}~\frac{\kappa}{g}~\frac{\mu}{z_+^2~A},\qquad \textrm{and}\qquad q_2~=~\sqrt{\frac{3}{2}}~\frac{\kappa}{g}~\frac{c~\mu}{A}.\]

The chemical potential is identified as limit $z\to 0$ of temporal component of gauge field \cite{Nakamura:2007nk,Sachan:2011iy}.  Since the chemical potential is related to asymptotic value of the gauge field, it is measured in the units of gauge fields. The scale, i.e. MeV, can be estimated by substituting the experimental results of meson masses in equations of motions as shown in \cite{Da_Rold:2005zs, Erlich:2005qh}.

The transition temperature vs chemical potential plots are shown in \autoref{fig:GB_tmu} (we have taken $N_f=2$ an $N_c=3$) for different values of Gauss-Bonnet coupling $\alpha$.  From the plots in  \autoref{fig:GB_free}, it is evident that the transition temperature (or more precisely deconfinement temperature) depends on the Gauss-Bonnet coupling  and the change is significant for larger values of the chemical potential. This has to do with the fact that the difference of free energies (eqn. \ref{freediff}) depend explicitly on Gauss-Bonnet coupling and the functions $F_i$ depend on $q$ and $q_2$, which in turn are directly proportional to the chemical potential. The increase in value of chemical potential increases the dependence  of free energy difference on Gauss-Bonnet coupling through the relation (\ref{freediff}), which results in significant change in transition temperature for higher values of chemical potential. In boundary theory, the dependence of transition temperature with respect to Gauss-Bonnet coupling is related to the corrections in terms of t-Hooft coupling.

All curves (for different values of Gauss-Bonnet coupling $\alpha$) meet at a single point (\autoref{fig:GB_tmu}) in both soft wall as well as hard wall models (the details of the  hard wall model are worked out in the appendix). At this point, the dependency of curve on Gauss-Bonnet coupling vanishes. We speculate that this point may correspond to change of the transition from first order to cross-over in QCD phase diagram.  It may be interesting to explore this point further and one may be interested in to get more insight about the crossover using dual theories.

\begin{figure}
  \begin{subfigure}[b]{0.5\linewidth} 
    \centering
    \includegraphics[width=8cm]{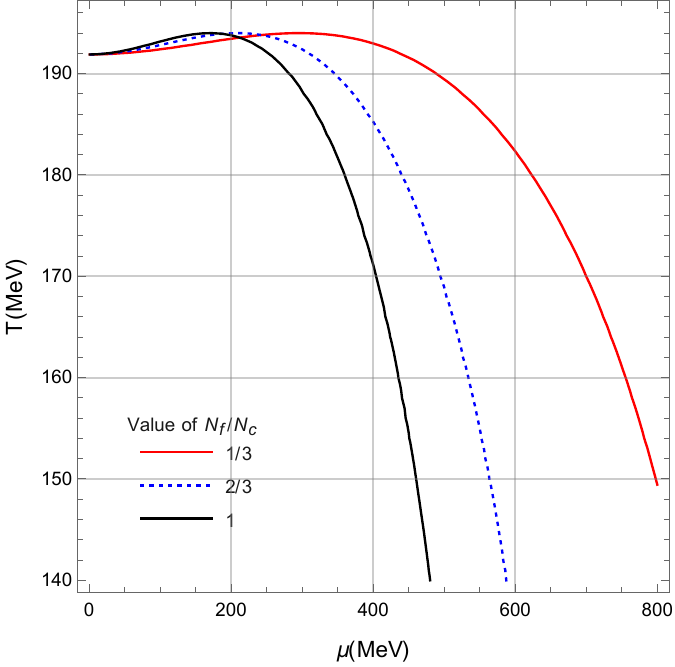}
    \caption{for $\alpha=-0.001$.}\label{fig:GB_tmua1}
  \end{subfigure}\hfill
  \begin{subfigure}[b]{0.5\linewidth}
    \centering
    \includegraphics[width=8cm]{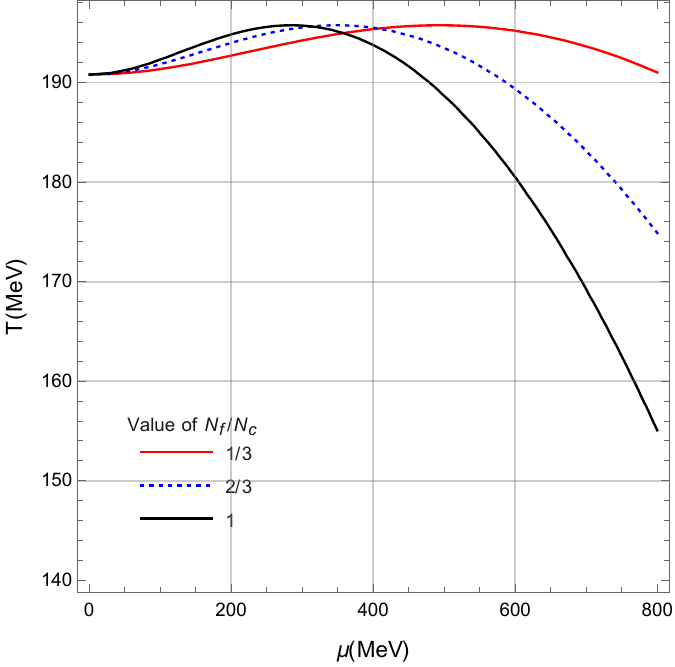}
    \caption{for $\alpha\to0$.}\label{fig:GB_tmua0}
  \end{subfigure}
  \caption{Transition temperature vs chemical potential for different values of  $N_f/N_c$ in soft wall model.}\label{fig:GB_tmun}
\end{figure}

In \autoref{fig:GB_tmun},  we have plotted transition temperature vs chemical potential for different values of $N_f/N_c$ for soft wall model. All these curves shows that the increasing value of $N_f/N_c$ ratio, shifts the curves towards the lower value of chemical potential. The plots in  \autoref{fig:GB_tmua0} shows the behavior of transition temperature for different values of $N_f/N_c$ in the limit $\alpha\to 0$. The variation of the ratio $N_f/N_c$ doesn't change the qualitatively features of $\mu-T$ plot.
 
\section{Conclusion \& Outlook}\label{sec:con}
\noindent
We have investigated the effects of Gauss-Bonnet coupling on the deconfinement transition in soft wall model of AdS/QCD. The Gauss-Bonnet terms are higher derivative corrections to Einstein gravity, and are related with string tension in holographic models. The variation of grand potential with respect to temperature is analyzed for different values of chemical potential. The introduction of Gauss-Bonnet coupling alters the transition temperature particularly for higher values of chemical potential.  We also obtained the results for the variation of transition temperature with respect to chemical potential for different values of $N_f/N_c$. We observed that the deconfinement temperature falls rapidly with respect to chemical potential as the value of $N_f/N_c$ is increased. 
 
We obtained the transition temperature versus chemical potential plot for the soft wall  model. There exists a point on $\mu - T $ plane which is almost fixed, i.e. independent of Gauss-Bonnet parameter. Interestingly, same conclusion is shown to be valid in hard wall model as well. This point may be a signature of cross over behavior as we approach from higher chemical potential to the lower one. In that case, our results are important as one can  pin-point the critical point which lies at the end of first order transition. In the absence of Gauss-Bonnet corrections, the holographic models only predict a first order phase transition. Our results in soft wall model of QCD seem to be in agreement with the values of transition temperature obtained in lattice models \cite{Steinbrecher:2018phh}. The predicted transition temperature in soft wall model is higher than that of the lattice models, but the range of chemical potential over crossover region agrees with that of lattice models. Thus, the predictions of soft wall model may not be exact but open a new way to examine the behavior of QCD near transition temperature.

Our analysis is valid for small values of Gauss-Bonnet coupling in soft wall model as we have analyzed our solutions up-to first order in the coupling parameter.  The numerical methods can be used to analyze the results for the general values of Gauss-Bonnet coupling. The comparison with lattice QCD results can be encouraging. Perhaps, a more rigorous framework is needed and we plan to study these observations for new type of black hole solutions and incorporating non-linear corrections to gauge part of the bulk action in the holographic models.
 
\section*{Acknowledgments:}
\noindent
S. Sachan is supported by CSIR-Senior Research
Fellowship, grant no. (09/013(0239)/2009-EMR-I). S. Siwach acknowledges the financial support from the DST Young Scientist project while a part of the work was completed.

\section*{Appendix}
\noindent
\section*{Hard wall model}
\noindent
We summarize the details of calculation for the Gauss-Bonnet corrected hard wall model in this section for the sake of completeness.  We begin with the Einstein-Maxwell action with Gauss-Bonnet term as given by the \autoref{action0}. After evaluating equation of motions and simplifying the various terms, the  action of (\ref{action0}) reduces to that of (\ref{action}). The principle difference between hard wall and soft wall model is the existence of dilaton field $\phi=cz^2$ in soft wall model. We can perform the calculations of hard wall model by substituting the  constant $c$ to zero. In addition we have to impose the infra-red (IR) cut-off $z_\mathrm{IR}$  in the hard wall model. This substitution leads to reduction of the action (\ref{action1}),
\be 
S~=~-\int~d^5x~\sqrt{g}~\left(\frac{-R+4\Lambda}{\kappa^2}\right),\nn
\ee
which is same as that of the action \ref{action}. 

Two study the Hawking-Page transition in hard wall model with Gauss-Bonnet corrections, we consider the line element of form,
\be 
ds^2~=~\frac{L^2}{z^2}\left(A^2
f(z)\,dt^2+\frac{dz^2}{f(z)}+\sum_{i=1}^{3}{dx^i}^2\right)
\ee
with  $A^2= \frac{1}{2} \left(\sqrt{1-8 \alpha }+1\right)$. 
\begin{eqnarray}
f(z)&=& \left\{\begin{array}{ll}
 \frac{1}{4\alpha}\left( 1-\sqrt{1-8\alpha(1-mz^4+q^2z^6)}\right) &~~~ \mathrm{for~ charged~ black~ hole}\\ \\
\frac{1}{4\alpha}\left( 1-\sqrt{1-8\alpha(1+q^2z^6)}\right)&~~~ \mathrm{for~ thermally ~charged ~AdS.}
\end{array}\right.
\end{eqnarray}
We simplify the action \ref{action} for above metric functions and regularize it using the procedure as mentioned in sections (\ref{sec:cb}) and (\ref{sec:tc}). The regularized action in hard wall model with Gauss-Bonnet corrections for charged black hole and for thermally charged AdS is written as,
\begin{align}
\frac{\Omega_1}{V_3}~&=~A \frac{\left(-1+\sqrt{1-8 \alpha }\right) \sqrt{1-8 \alpha }   \left(q^2 z_+^6+1\right)}{2 \left(8 \alpha +\sqrt{1-8 \alpha }-1\right) \kappa ^2 z_+^4}\nn\\
\frac{\Omega_2}{V_3}~&=~-A\frac{8 \alpha  \left(3 \sqrt{1-8 \alpha  \left(q^2\, z_\mathrm{IR}^6+1\right)}-2 q^2 \,z_{IR}^6-5\right)-5 \sqrt{1-8 \alpha  \left(q^2\, z_\mathrm{IR}^6+1\right)}+5}{4~ \alpha ~ \kappa ^2~ z_\mathrm{IR}^4 ~\sqrt{1-8 \alpha  \left(q^2\, z_\mathrm{IR}^6+1\right)}},\nn
\end{align}
respectively.  In above equation, we have used values of $A,$ and  $z_+$ as in soft wall model, value of $q$ for charged black hole as $\sqrt{\frac{2}{3}}~\frac{\kappa}{g}~\frac{\mu}{z_+^2~A}$, for thermally charged AdS as $\sqrt{\frac{3}{2}}~\frac{\kappa}{g}~\frac{\mu}{z_\mathrm{IR}^2~A}$. The integrals in the hard wall model have a upper limit given by the value of IR cutoff $z_{IR}$. The value of IR cut-off is fixed by matching the specra of mesons and is given by, $z_{IR} =1/(323~ MEV)$ \cite{Erlich:2005qh}. The transition temperature vs chemical potential plots are obtained by equating,
\be
\frac{\Delta\Omega}{V_3}~=~\frac{\Omega_1-\Omega_2}{V_3},\nn
\ee
to zero and are shown in Fig. 3(b).  In the main text, we have compared our results of the soft wall model with hard wall model using the above formalism.

\bibliographystyle{elsarticle-num}
\bibliography{<your-bib-database>}


\end{CJK*}
\end{document}